\documentstyle[psfig,conf_iap,10pt]{article}
\def\kms{km~s$^{-1}$~}

\begin{document}
\heading{Spiral Galaxies and the Peculiar Velocity Field} 
\par\medskip\noindent
\author{Riccardo Giovanelli$^1$}
\address{Dept. of Astronomy, Cornell University, Ithaca, NY 14853, USA}
\begin{abstract}
Monopole and dipole signatures of the peculiar velocity field as derived from
the SFI sample of field spirals and the SCI and SC2 samples of cluster spirals
are presented. The monopole exhibits no evidence of a `Hubble bubble'
within 7000 \kms as suggested by \cite{Z98}; the dipole of the reflex motion
of the Local Group converges to the CMB dipole within less than 6000 \kms,
and remains consistent with it, when referred to the reference frame 
constituted by the SC2 cluster sample, which extends to 20,000 \kms.
\end{abstract}
\section{The SCI, SFI and SC2 Surveys}

We report on the results of peculiar velocity measurements obtained from 
three samples of spiral galaxies, based on the Tully--Fisher technique which
combines I--band photometry and either 21 cm or H$\alpha$ long--slit 
spectroscopy. The three samples are: (a) SCI, which includes 782 galaxies
in the fields of 24 clusters within $cz\sim 9000$ \kms (\cite{G97a},
\cite{G97b}); (b) SFI, which includes 1631 field galaxies out to $cz\simeq 
9000$ \kms (\cite{H98a},\cite{H98b}); and (c) SC2, which includes 522 galaxies 
in the fields of 52 clusters out to $cz\simeq 20,000$ \kms(\cite{D97},
\cite{D98a}).

\section{No Hubble Bubble}

Recently, it has been suggested by \cite{Z98} that the volume within 
$cz\simeq 7000$ \kms is subject to a Hubble acceleration of $6.6\pm2.2$\%,
resulting from a local underdensity of 20\%, surrounded by an overdense shell.
The result is based on the distances to 44 SN of type Ia. In Figure 1,
we test that result by using the combined sample of 76 clusters of SCI plus
SC2. The distances to each individual cluster have typical accuracy of 
5\%, and the cluster sample is well distributed over the sky. Figure 1 displays 
$\delta H/H = V_{pec}/cz_{tf}$ against $hd = cz_{tf}/100$, where $V_{pec}$ is 
the peculiar velocity in the CMB reference frame and $cz_{tf}$ is the TF 
distance in Mpc, in the same frame. Unfilled symbols represent clusters with 
poor distance determinations, based on fewer than 5 galaxies with TF 
measurements per cluster. Fig. 1 
illustrates how, at small distances, the deviations from Hubble flow are
dominated by the motions of nearby groups, comparable in amplitude to those
of the LG (611 \kms). At distances larger than 35$h^{-1}$ Mpc, the monopole
of the cluster peculiar velocity field exhibits no significant change of
value, anywhere up to 200 Mpc. The Hubble bubble suggestion of \cite{Z98}
is not corroborated by this larger and more accurate data set. 

\begin{figure}
\centerline{\vbox{\psfig{figure=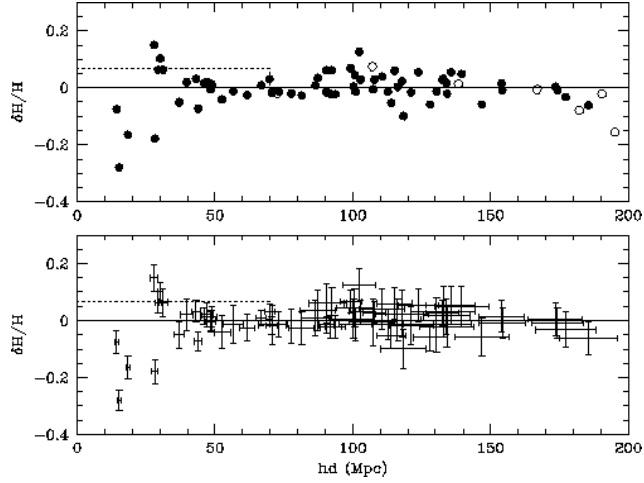,height=6.3cm}}}
\caption[]{No Hubble Bubble in the local universe. $\delta H/H = 0$ line
identifies null deviation from Hubble flow. The dotted line is the effect
of the Hubble bubble proposed by \cite{Z98}. The lower panel
shows the error bars associated with the filled data points shown above.}
\end{figure}

\section{Dipoles of the Field Spiral Sample (SFI)}

Figure 2 displays the dipoles of the reflex motion of the Local Group (LG)
with respect to field galaxies in shells 2000 \kms thick. The dashed
line in panel 2a corresponds to the amplitude of the CMB dipole, of 611
\kms. The three sets of symbols identify different ways of computing the
peculiar velocities, using different subsets of the data or adopting a
direct (stars) or inverse TF relation. The SFI sample indicates convergence
to the CMB dipole, in the motion of the LG with respect to galaxies 
in shells, when a radius of a few thousand \kms is reached. Convergence 
is achieved both in amplitude and apex direction, The  dipole of \cite{LP94}
is excluded with a high level of confidence. See \cite{G98a}
for further details.
\begin{figure}
\centerline{\vbox{\psfig{figure=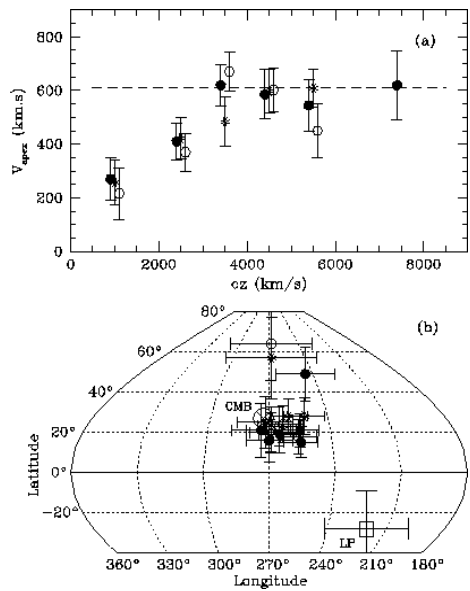,height=7.7cm}}}
\caption[]{Amplitudes (a) and apex directions (b) of the peculiar velocity
dipole for SFI. `CMB' and `LP' identify
the apices of the CMB and \cite{LP94} dipoles. The dashed line in
(a) is the amplitude of the CMB dipole of 611 \kms.}
\end{figure}

\section{Dipoles of the Cluster Samples (SCI and SC2)}

The dipole moments of the reflex motion of the LG with respect to (i) the
subset of clusters farther than 4000 \kms in the SCI cluster sample and 
(ii) the SC2 sample both coincide, within the errors, with that of the CMB.
This coincidence is matched both in amplitude and apex direction. Figure 3
displays the error clouds for the coordinates of the apices of data sets 
derived from the SC2 
sample, plotted against supergalactic Cartesian coordinates. One and two--sigma
contours are plotted, as well as the locations of the CMB and the Lauer \&
Postman dipoles. The latter can be excluded by our data with a high level
of confidence. The bulk flow within a sphere of 6000 \kms radius, as derived 
from SCI, is between 140 and 320 \kms, in the CMB reference frame, somewhat
lower than obtained by other techniques (e.g. \cite{daC96},\cite{D94}). See 
\cite{D98b} and \cite{G98b} for details.

\section {Conclusion}

The reflex motion of the LG converges to the CMB dipole amplitude and direction 
within 6000 \kms. No evidence is found for a local Hubble bubble.

\acknowledgements{The results presented here were obtained by the combined
efforts of M. Haynes, D. Dale, E. Hardy, L. Campusano, M. Scodeggio, J. Salzer, 
G. Wegner, L. da Costa, W. Freudling and the author. They are based on 
observations that were carried out at several observatories, including Palomar, 
Arecibo, NOAO, NRAO, Nan\c cay, MPI and MDM. NRAO, NOAO and NAIC are operated 
under management agreements with the National Science Foundation respectively by
AUI, AURA and Cornell University. The Palomar Observatory is operated by Caltech 
under a management agreement with Cornell University and JPL. This work was 
supported by NSF grants AST94--20505 and AST96--17069.}

\begin{figure}
\centerline{\vbox{\psfig{figure=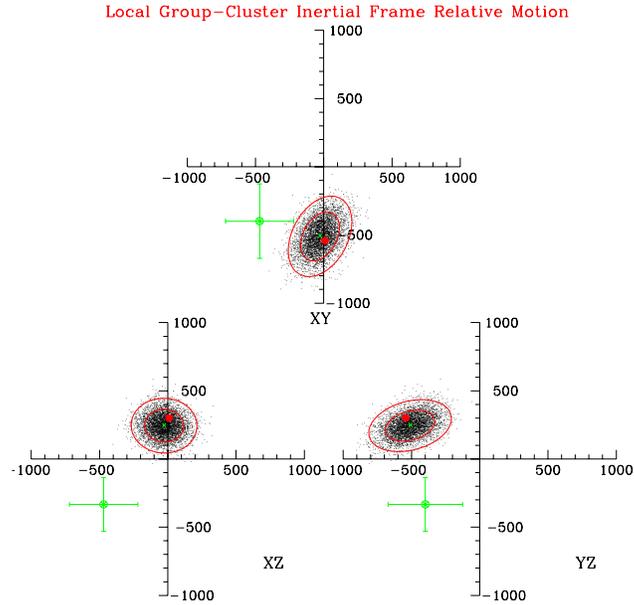,height=8.5cm}}}
\caption[]{Error clouds of the dipole of the reflex motion of the LG
with respect to SC2. The error clouds (with 1--$\sigma$
and 2--$\sigma$ contours) are plotted in supergalactic Cartesian
coordinates. The CMB dipole is a filled symbol, which appears enclosed, in all 
plots, within the 1--$\sigma$ contours; the large cross is the
dipole of \cite{LP94}, which is excluded by the data.}
\end{figure}
\vfill
\end{document}